\documentclass[fleqn,twoside]{article}
\usepackage{espcrc2}

\usepackage{graphicx}
\usepackage[figuresright]{rotating}

\newcommand{\AmS}{{\protect\the\textfont2
  A\kern-.1667em\lower.5ex\hbox{M}\kern-.125emS}}

\title{THE POMERON IN EXCLUSIVE $J/\psi$ VECTOR MESON PRODUCTION}

\author{R.~Fiore\address{Dipartimento di Fisica, Universit\`a della Calabria \\
                Instituto Nazionale di Fisica Nucleare, Gruppo collegato di Cosenza \\
                I-87036 Arcavata di Rende, Cosenza, Italy}
        \thanks{FIORE@CS.INFN.IT},
       L.L.~Jenkovszky\address{Bogolyubov Institute for Theoretical Physics \\
                       Academy of Science of Ukraine \\
                       UA-03143 Kiev, Ukraine}
                      \thanks{JENK@GLUK.ORG},
       F.~Paccanoni\address{Dipartimento di Fisica, Universit\`a di Padova\\
                    Instituto Nazionale di Fisica Nucleare, Sezione di Padova\\
                    Via F. Marzolo 8, I-35131 Padova, Italy}
                   \thanks{PACCANONI@PD.INFN.IT} and 
       \underline{A.~Prokudin}\address{Dipartimento di Fisica Teorica, Universit\`a
Degli Studi di Torino \\
                  Instituto Nazionale di Fisica Nucleare, Sezione di Torino \\
                  Via P. Giuria 1, I-10125 Torino, Italy \\
                  Institute For High Energy Physics, 142284 Protvino, Russia}
                            \thanks{PROKUDIN@TO.INFN.IT}
                            \thanks{Talk presented by A. Prokudin at Photon 2003, Frascati, Italy, 7-11 April 2003}
}
       
\begin{document}

\begin{abstract}
An earlier developed model for vector meson photoproduction,
based on a dipole Pomeron exchange, is extended to electroproduction.
Universality of the non linear Pomeron trajectory is tested by fitting
the model to ZEUS and H1 data as well as to CDF data on $\bar pp$ elastic
scattering.
\vspace{1pc}
\end{abstract}

\maketitle

\section{Introduction}

Elastic production of vector mesons in electron-proton interaction
has provided a deeper understanding of the diffraction phenomenon
and finds a sensible description in a variety of models. 
The first attempts to describe diffractive photoproduction were based
on vector dominance model~\cite{SC} and Regge theory~\cite{DL}.
Since
various aspects of the deep inelastic scattering and of elastic
processes are both present in photoproduction it is quite
natural that perturbative QCD can help to understand many features
of the HERA experimental results. Examples where perturbative QCD
has been applied to this process can be found in 
Ref.~\cite{RY}. 

In these perturbative
calculations regarding diffractive processes, that in this case
have characteristic features of the elastic ones, non perturbative
contributions are present and their description becomes an
important ingredient of the theoretical model while lying outside
perturbation theory. Hence, models based on Regge pole
phenomenology maintain their important task in helping to
construct a representation of non perturbative aspects of the
scattering amplitude. For the processes under consideration many
papers based on Regge poles exchange successfully
reproduced~\cite{JMP,FJP} new experimental HERA data.

The aim of this paper is to expound the properties of the
most important and intricate Regge exchange: the vacuum, or
Pomeron exchange. $J/\psi$ photoproduction and, apart from small
subleading contributions, $\phi(1020)$ photoproduction are genuine
Pomeron filters that, together with very high energies reactions,
permit a careful study of the non perturbative features of
diffraction. 

\section{The model}

A convenient way to obtain rising cross sections with the Pomeron
intercept equal to one assumes that the Pomeron is a double Regge
pole. This means that the $t$-channel partial wave, corresponding to
the Pomeron exchange, has a double pole for $\ell=\alpha(t)$. In
this choice we are comforted by the numerous successes of this
model in its applications to hadronic reactions~\cite{JMP,FJP,LJ,[8]}.

We choose the invariant scattering amplitude in
the form
\begin{equation}
\displaystyle A(s,t)=i f(t)
\left(-i\frac{s}{s_0}\right)^{\alpha_P(t)} 
\left[\ln\left(-i\frac{s}{s_0}
\right)+g(t)\right]\;, \label{d1}
\end{equation}
where $g(t)$ and $f(t)$ are functions, for the moment
undetermined, of the momentum transfer. $\alpha_P(t)$ is the
Pomeron trajectory with $\alpha_P(0)=1$.

The choice
of the function $f(t)$, that represents the product of the
vertices $\gamma$-Pomeron-meson and proton-Pomeron-proton,
 will be made by
imposing the condition that the Pomeron exchange is pure spin
$\alpha_P$ exchange. It has been shown in~\cite{FP} that this
constraint leads to a vertex of the form $[(\alpha_P(t)-1)f_1(t)
+(\alpha_P(t)+1)f_2(t)]$ and, in the neighborhood of $t=0$, to a
term vanishing with $t$ whatever the form of the trajectory could
be. 

 It has been shown in Ref.~\cite{FJP} that the simple form for the
elastic differential cross section of vector meson photoproduction

\begin{eqnarray}
\nonumber
\frac{d\sigma}{dt}=4\pi\left[a\,e^{bt}+ct\,e^{dt}\right]^2\left(
\frac{s}{s_0}\right)^{2\alpha_P(t)-2} \cdot \\ 
\cdot \left[\left(\ln\frac{s}{s_0}+g\right)^2+
\frac{\pi^2}{4}\right]\;, 
\label{d2}
\end{eqnarray}

where $g$ is a constant, gives a good quality fit to the
experimental data~\cite{ZEUS,H1}. We notice that the form
(\ref{d2}) satisfies also the aforesaid conditions and will be
adopted in this paper. 

Since the amplitude, in the Regge
form, should have no essential singularity at infinity in the cut
plane, ${\cal R}e\:\alpha(s)$ is bounded by a constant, for $s\to
\infty$, and this leads to the bound
$|\alpha(s)|< Ms^q$ for  $s\to\infty$ 
with $q<1$ and $M$ an arbitrary constant. The
choice~\cite{RF,FJP}
\begin{equation}
\alpha_P(t)=1+\gamma(\sqrt{t_0}-\sqrt{t_0-t}), \label{d3}
\end{equation}
where $t_0=4m_{\pi}^2$ and $\gamma=m_{\pi}/1\,GeV^2$, satisfies the
above conditions and reproduces the standard Pomeron slope at
$t=0$, $\alpha'_P(0)\simeq 0.25\;GeV^{-2}$. Eq.~(\ref{d3}) for the
trajectory defines uniquely the model for photoproduction. 

Consider now electroproduction of a vector meson. As
noticed in Refs.~\cite{ZEUS1,H11} a commonly adopted form for the $Q^2$
dependence of the $J/\psi$ cross section is
\begin{equation}
\sigma_{tot}^{\gamma^*\,p\;\to J/\psi\,p} \propto
\frac{1}{(1+Q^2/M^2_{J/ \psi})^n}\;\; , \label{d4}
\end{equation}
where $n\sim 1.75$, according to the ZEUS Collaboration~\cite{ZEUS1}, and $n\sim
2.38$ according to the H1 Collaboration~\cite{H1}. 

For large $Q^2$ all the
amplitudes but the double flip one, for diffractive vector meson
electroproduction, can be evaluated in perturbative QCD
~\cite{KNZ}. In the longitudinal photon amplitudes, a factor
$Q/M_{J/\psi}$ is a consequence of gauge invariance irrespective
of the detailed production dynamics. If we consider only the
dominant twist $s$-channel helicity conserving amplitudes, the
factor in Eq.~(\ref{d4}) thus finds a natural explanation. The
$Q^2$ dependence, however, will appear also in the strong coupling
and in the gluon structure function through the hard scale of
perturbative QCD~\cite{KNZ}. 

In our
approach, based on Regge pole theory, the factor (\ref{d4}) will
be certainly present in electroproduction, multiplying the
differential cross section (\ref{d2}), but this will not complete
all the possible corrections. Since, in the dipole Pomeron
formalism, the product of the vertices can affect the parameter
$g$, all the parameters can acquire a weak $Q^2$ dependence. We
neglect this dependence in $a,\,b,\,c,\,d $ and assume that $g$
varies as $g\times [1+Q^2/(Q^2+M_V^2)]^{\gamma}$ where $\gamma$,
if this assumption is correct, is small. One can interpret this
functional dependence of $g$ as coming from a $Q^2$ dependence of
$s_0$ in $\ln(s/s_0)$.  \vskip 0.3cm The final form of the
differential cross section is:
\begin{eqnarray}
\nonumber 
\frac{d\sigma}{dt}= 4\pi\left (1+\frac{Q^2}{M^2_{J/\psi}}
\right)^{-\beta} \left [ a\,e^{bt}+ct\,e^{dt}\right]^2 \cdot \\
\left(\frac{s}{s_0} \right)^{2\alpha_P (t)-2} 
\left(\left[\ln\left(\frac{s}{s_0}\right)+g(Q^2)\right]^2+ \frac{\pi^2}{4}\right)\;, \label{d5}
\end{eqnarray}

where, for $Q^2=0$, all the parameters have the same value as
for photoproduction. We notice that the value of $\beta$ includes a factor
$(1+Q^2/M^2_{J/\psi})$ that comes from the contribution of the longitudinal
 amplitude, relevant at $Q^2\neq 0$, which leads to
$|A|^2=|A_T|^2+|A_L|^2$. The approximate relation 
$A_L \sim Q\;A_T/M_{J/\psi}$ can be applied in this phenomenological approach. 

In the following Section
the parameterizations (\ref{d2}) and (\ref{d5}) will be applied to $J/\psi$
photoproduction and electroproduction. 

\section{$J/\psi$ photoproduction and electroproduction}
Following the analysis of Ref.~\cite{FJP} we apply Eq.~(\ref{d2})  to the new dataset
of $J/\psi$ photoproduction \cite{ZEUS}\footnote{The data are available from \cite{data}.}
\begin{equation}
\gamma + p \rightarrow J/\psi + p
\label{z6}
\end{equation}
In the experiment, the $J/\psi$ is identified from its leptonic decay 
modes, electron $J/\psi\rightarrow e^+e^-$ or muon $J/\psi\rightarrow \mu^+\mu^-$ pair, with different systematic errors specific
to the electron  or muon  decay channel. For this reason the dataset \cite{ZEUS}
presents two separate measurements of the process Eq.(\ref{z6}), according 
to the way of the $J/\psi$ detection. Hence, as a first attempt,
we limit our fit to the region $W \leq 160\;{\rm GeV}$, where data from both
decay channels are given, and check the predictions of the model for
the differential cross section and the total integrated cross section.

As noticed in the previous paper \cite{FJP}, the parameter $d$ varies little 
in the fit, so that we keep the same value $d=0.851\; {\rm GeV^{-2}}$  
fixed, thus leaving only four parameters free. As in previous
paper~\cite{FJP} we set $s_0=1$ GeV$^2$. In order to avoid the
region of inelastic background we limit the $t$ region to $|t|<1\; {\rm  GeV^2}$.
In the fit we use differential cross sections only.
For the electron channel we have obtained the results shown
in Column 2 of Table~\ref{Table 1.}, with $\chi^2/{\rm d.o.f.}=1.5$.
For the muon channel the results are presented in Column 3 of 
Table~\ref{Table 1.}, with $\chi^2/{\rm d.o.f.}=1.0$.

{\small
\begin{table*}[htb]
\begin{center}
\begin{tabular}{|l|l|l|l|}
\hline
\multicolumn{4}{c}{Photoproduction} \\
\hline
 & $e^+e^-$ channel & 
   $\mu^+\mu^-$ channel & 
   $e^+e^-$ channel  
   \\
 & $W<160\; {\rm GeV}$ & 
   $W<160\; {\rm GeV}$ & 
   $W<300\; {\rm GeV}$  
   \\
\hline
a $[{\rm GeV}^{-2}]$ & 
  $(1.8 \pm 0.1)\cdot 10^{-3}$ &
  $(1.83 \pm 0.09)\cdot 10^{-3}$ &
  $(1.97 \pm 0.13)\cdot 10^{-3}$ 
  \\
b $[{\rm GeV}^{-2}]$ & 
  $1.55   \pm    0.49$ &
  $2.25   \pm    0.24$ &
  $1.40   \pm    0.51$ 
  \\
c $[{\rm GeV}^{-4}]$ & 
  $(-0.48 \pm 0.54)\cdot 10^{-3}$ &
  $(-0.97 \pm 0.19)\cdot 10^{-3}$ &
  $(-0.35 \pm 0.67)\cdot 10^{-3}$
 \\
d $[{\rm GeV}^{-2}]$ & 
  $0.851$ &
  $0.851$ &
  $0.851$  
  \\
g  & 
  $-4.23  \pm     0.37$ &
  $-4.25  \pm     0.22$ &
  $-4.58  \pm     0.29$ 
\\
\hline
%
\end{tabular}
\end{center}
\caption{ Values of parameters obtained by fitting
$J/\psi$ photoproduction data.  
\label{Table 1.}}
\end{table*}}
If we use all the two channel data altogether we obtain a very
high $\chi^2/{\rm d.o.f.} = 1.9$. 
To implement a better analysis one needs a more complete set
of data on differential cross sections in both channels.

In order to proceed with the fitting procedure we must
choose one of these channels. As soon as our model is valid for high
energies and the data on $J/\psi$ exclusive photoproduction
in $e+e-$ channel cover a region of higher energies ($30<W<300$ GeV)
and has a better statistic
than $\mu+\mu-$ channel ($30<W<160$ GeV), we choose the $e+e-$ channel
data. The result is presented in Column 4 of 
Table~\ref{Table 1.}, with $\chi^2/{\rm d.o.f.}=1.2$.

Without any fitting we achieve a good agreement with the data
on integrated elastic cross section, $\chi^2/{\rm point} = 0.95$. 
The high error of $c$ is due to the scarcity of the data on
the differential cross section in the region $0<|t|<1\; {\rm GeV^2}$.
A more complete set of high accurate data will allow us
to arrive at a definite conclusion about the values
of parameters.  

\begin{figure}[htb]
\includegraphics*[scale=0.35]{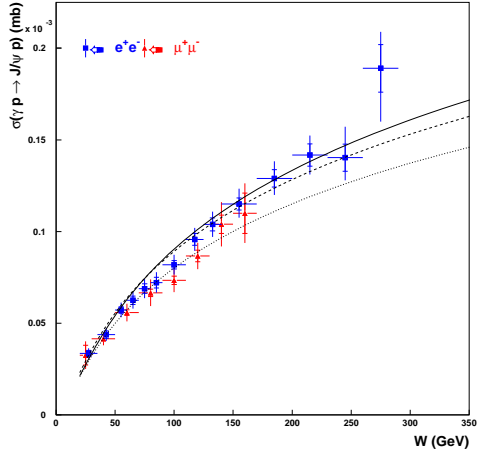}
\vskip -1.2cm
\caption{Elastic
cross section of $J/\psi$ photoproduction.
The dashed line corresponds to $J/\psi \rightarrow e^+e^-$ channel fit
(Column 2 of Table \ref{Table 1.}). The dotted line corresponds to 
$J/\psi \rightarrow \mu^+\mu^-$ channel fit
(Column 3 of Table \ref{Table 1.}). The solid line corresponds to 
$J/\psi \rightarrow e^+e^-$ channel fit
(Column 4 of Table \ref{Table 1.}).}
\label{fig:jpsi}
\end{figure}

\begin{figure}[htb]
\includegraphics*[scale=0.35]{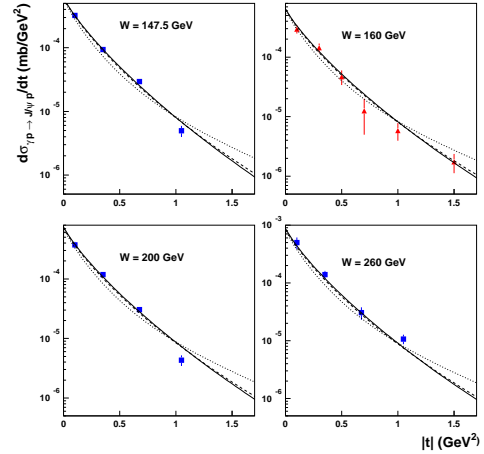}
\vskip -1.2cm
\caption{Differential cross section of
exclusive $J/\psi$ photoproduction
for 147.5$\le W\le$ 260 GeV. Line aliases and symbols are the same as
in Fig. \ref{fig:jpsi}.}
\label{fig:jpsid2}
\end{figure}

Now we use Eq. (\ref{d5}) in order to describe electroproduction
of $J/\psi$. We fix all the parameters obtained by fitting
the photoproduction data (see Column 4 of Table \ref{Table 1.}.) and
fit two parameters $\beta$ and $\gamma$ to the dataset \footnote{The data are available from \cite{data}, \cite{data1}.}.
The values of the parameters are the following:
$\beta = 1.94 \pm 0.42$, $\gamma = 0.69   \pm   0.24$ and
$\chi^2/{\rm d.o.f.} = 0.81 $.

The factor $[1+Q^2/(Q^2+M_V^2)]^{\gamma}$ grows up to
1.5 in the available region of photon virtuality $0<Q^2<50\; {\rm GeV^2}$.

In the case of $\gamma = 0$ we obtain $\beta=2.86   \pm    0.09$
and $\chi^2/{\rm d.o.f.}=1.07$. We proved that the fit is rather 
insensible to the value of $0<\gamma<1$.

\begin{figure}[htb]
\includegraphics*[scale=0.35]{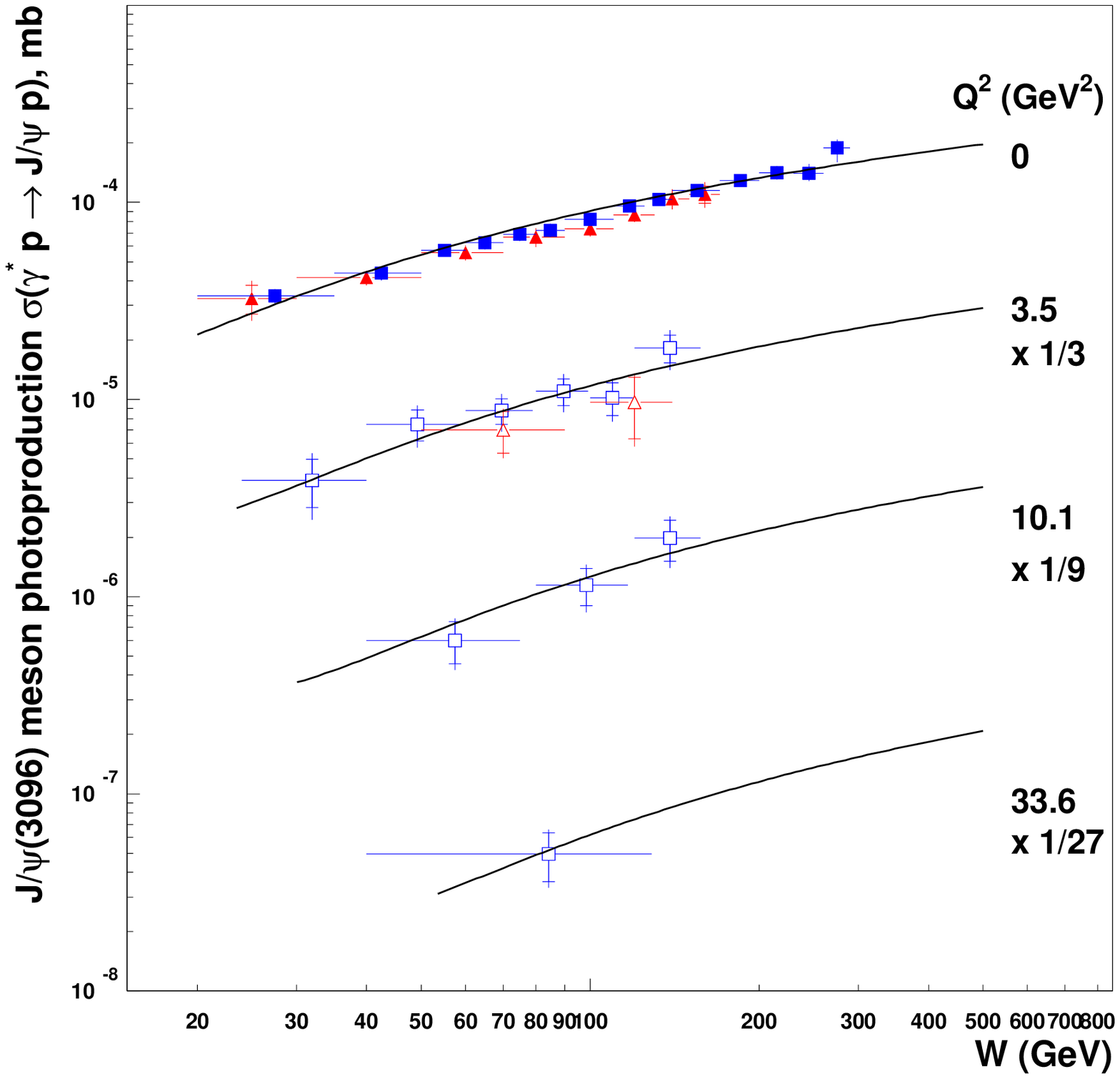}
\vskip -1.2cm
\caption{Cross section of
exclusive $J/\psi$ electroproduction 
as a function of $W$. }
\label{fig:jpsiq}
\end{figure}

\section{Pomeron universality}

The model we consider is consistent with $s$-channel unitarity and asymptotic
factorizability. Universality, in this context, refers to the choice
(\ref{d3}) for the Pomeron trajectory that provides a reliable description
of exclusive vector meson production. The conjecture that the trajectory
in Eq. (\ref{d3}) is universal is supported by the following example.

We consider the proton-antiproton scattering at sufficiently high energies,
where only the Pomeron presumably contributes. Following tradition \cite{DL1},
it is a customary practice to adopt a linear Pomeron trajectory in order 
to describe hadronic interactions. In a different approach \cite{LJ,JJS}
that provides a satisfactory fit to $pp$ and $\bar pp$ data a square root 
trajectory similar to  that of Eq.~(\ref{d3}) has been preferred. It is 
interesting to update this last fit using Eq.~(\ref{d3}) and the same 
parameters adopted for photoproduction: 
$t_0=4 m_\pi^2$ and $\gamma=m_\pi/1 {\rm GeV^2}$.

In order to use the asymptotic
formula, we choose the data on the differential cross section at energies
$\sqrt{s}=546$ GeV and $1.8$ TeV \cite{CDF}. As we take into account 
neither Pomeron daughters nor possible odderon contributions,
we concentrate on the region of low $|t|$, $0<|t|<0.2 \; {\rm GeV^2}$. The
result is presented in Table~\ref{Table 3.}, with  
$\chi^2/{\rm d.o.f.} = 1.04 $.
{\small
\begin{table}[htb]
\begin{center}
\begin{tabular}{ll}
a = & $0.41 \pm 0.01\; [{\rm GeV}^{-2}]$, \\
b = & $7.61   \pm    3.36\; [{\rm GeV}^{-2}]$, \\
c = & $-1.12 \pm 1.40 [{\rm GeV}^{-4}]$, \\
d = & $7.72\pm 0.52\; [{\rm GeV}^{-2}]$, \\
g = & $2.86 \pm 0.41 $. \\
\end{tabular}
\end{center}
\caption{ Values of parameters obtained by fitting
$p\bar p$ data.   
\label{Table 3.}}
\end{table}}

\begin{figure}[htb]
\includegraphics*[scale=0.35]{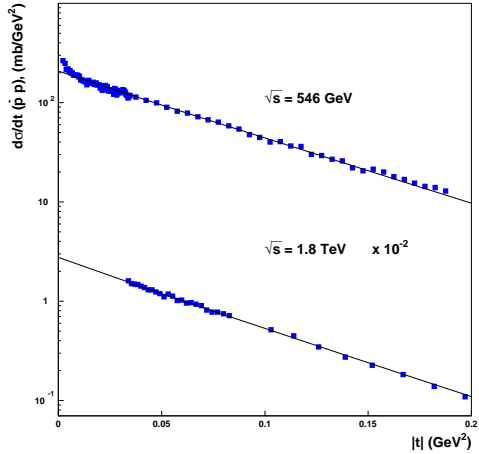}
\vskip -1.2cm
\caption{Differential cross section of
elastic $\bar pp$ scattering at the energies
$\sqrt{s}=546$ GeV and $1.8$ TeV. }
\label{fig:pbarp}
\end{figure}

\begin{figure}[htb]
\includegraphics*[scale=0.35]{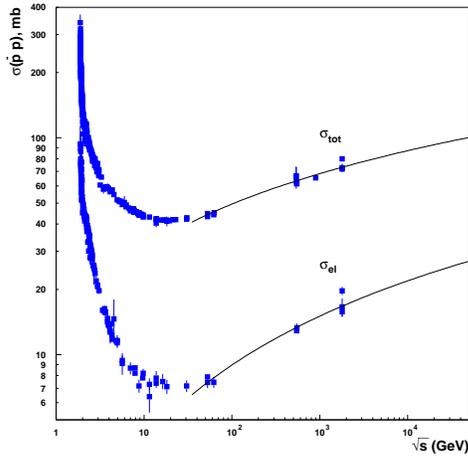}
\vskip -1.2cm
\caption{Elastic and total cross sections of
$\bar pp$ scattering.}
\label{fig:pbarpel}
\end{figure}

In Figs.~\ref{fig:pbarp},~\ref{fig:pbarpel} we depict respectively the 
results of the fit for the differential cross section and the predicted
total and elastic cross sections of $\bar pp$ scattering.

The non linear trajectory of Eq.~(\ref{d3}) provides a satisfactory agreement
with the data also for this hadronic process.
We consider the obtained result as an argument in support of the Pomeron
universality.

\section{Conclusions}

The aim of this paper was to study the Pomeron exchange in reactions where
non leading contributions are absent or negligible. We have chosen $J/\psi$
and $\phi(1020)$ photoproduction and electroproduction as Pomeron filters.

Our analysis is based on the dipole Pomeron model assuming a Pomeron trajectory with 
intercept equal to one and a non linear $t$-dependence. The choice of the 
vertices is based on covariant Reggeization as explained in Section 2 of
Ref. \cite{FP}. To reduce the number of free parameters we have used 
an approximate form of the vertex. As a result, we have obtained a 
good description of the data on $J/\psi$ and $\phi(1020)$ (see details
in Ref.~\cite{ourpaper})
photoproduction and electroproduction.

To demonstrate the universality of the chosen trajectory we applied the model
to $\bar pp$ scattering at sufficiently high energies where only the Pomeron
contributes.The good agreement with the experimental data is an argument 
in favor of the chosen Pomeron trajectory.

We are convinced to have reached a deeper understanding of the properties of
the soft dipole Pomeron.

\section*{Acknowledgments}

We would like to thank 
Ilya Ginzburg, Martin Block and Alessandro Papa
for helpful comments and Michele Arneodo and Alessia Bruni for fruitful discussions 
on the ZEUS data.

\end{document}